\documentstyle[amssymb,prl,aps,epsfig]{revtex}

\begin{document}

\title{Granular superconductivity in the cuprates evinced by finite size effects in
the specific heat and London penetration depths}
\author{T. Schneider}
\address{Physik-Institut der Universit\"{a}t Z\"{u}rich, Winterthurerstrasse 190,\\
CH-8057 Z\"{u}rich, Switzerland}
\maketitle

\begin{abstract}
We review and refine the finite size scaling analysis of specific
heat and London penetration depths data of cuprate superconductors
and compare it to the analysis of specific heat measurements near
the superfluid transition of $^{4}$He confined to 1$\mu $m$^{3}$
cylindrical boxes. This system crosses from $3D$ to $0D$ behavior
near the transition. This has a marked effect on the specific heat
as seen by a pronounced rounding of the maximum and a shift to a
temperature lower than the transition temperature of the bulk
system. The region in between the $3D$ to $0D$ crossover uncovers
the contributions from the surface and the edges of the
cylindrical boxes. Our finite size scaling analysis of the
specific heat and London penetration depth of high quality
YBa$_{2}$Cu$_{3}$O$_{7-\delta }$ and Bi$_{2}$Sr$_{2}$
CaCu$_{2}$O$_{8+\delta }$ single crystals uncover essentially the
same crossover phenomena, including evidence for surface and edge
contributions. This implies that the bulk samples break into
nearly homogeneous superconducting grains of rather unique extent
and with that granular superconductivity.
\end{abstract}
\bigskip
\begin{center}
To appear in the proceedings of \ Symmetry and Heterogeneity in
High Temperature Superconductors, Erice-Sicily: 4-10 October 2003
\end{center}

Since the discovery of superconductivity in cuprates by Bednorz
and M\"{u}ller\cite{bed} a tremendous amount of work has been
devoted to their characterization. The issues of inhomogeneities,
granularity, and their characterization are essential for several
reasons, including: (I) If inhomogeneity is an intrinsic property,
a re-interpretation of experiments, measuring an average of the
electronic properties, is unavoidable. (II) Inhomogeneity may
point to a microscopic phase separation, i.e. superconducting
grains, embedded in a non-superconducting matrix. (III) There is
neutron spectroscopic evidence for nanoscale cluster formation and
percolative superconductivity in various
cuprates\cite{mesot,furrer}. (IV) Nanoscale spatial variations in
the electronic characteristics have been observed on the surface
of Bi$_{2}$Sr$_{2}$CaCu$_{2}$O$_{8+\delta }$ with scanning
tunnelling microscopy (STM)\cite{liu,chang,cren,lang}. They reveal
a spatial segregation of the electronic structure into 3nm
diameter superconducting domains in an electronically distinct
background. (V) Remarkably, in the experimentally accessible
temperature regime, the critical properties of the superconductor
to normal state transition are found to fall into the 3D-XY
universality class, except from a rounding of the transition\cite
{tsda,ffh,tshkws,ohl,hub,jacc,kamal,kamal2,pasler,tsjh,tsjs,jhts,jhts2,tsjshou,book,meingast,tsphysB,osborn}.
This, however, is the result expected according to the Harris
criterion \cite{harris}, which states that short-range correlated
and uncorrelated disorder is irrelevant at the unperturbed
critical point, provided that the specific heat exponent $\alpha $
is negative. Since in the 3D-XY universality class $\alpha $ is
negative\cite{peliasetto}, the rounding of the transition appears
then to uncover a finite size effect\cite {cardy,privman}. Indeed,
consistency of the rounded transition in the specific heat and
penetration depths with a finite size effect has been established
in a variety of cuprate superconductors. The finite size scaling
analysis revealed that homogeneity is restricted to grains with
nanoscale extension\cite{book,tsrkhk,varenna,tsdc,bled,rktshkp}.
(VI) On the other hand, single crystals of cuprate superconductors
conduct in the normal state along any direction. Accordingly, bulk
superconductivity requires the grains to percolate. (VII) If this
finite size scenario holds true, the finite size behavior of the
superconductor to normal state transition should resemble features
of the superfluid transition of confined $^{4}$He, which falls
into the 3D-XY universality class, as well
\cite{coleman,kahn,murphy,kimball}. In particular, the finite size
effect should also provide information on the geometry of the
grains, e.g. their surface, edges, corners, etc.\ (VIII) Noting
that in Y$_{1-x}$Pr$_{x}$Cu$_{3}$O$_{7-\delta }$\cite{tsrkhk} and
YBa$_{2}$Cu$_{4}$O$_{8}$\cite{rktshkp} the transition temperature
increases with reduced extent of the grains along the c-axis, the
geometry turns out to be a relevant feature. Indeed, if the grain
is in contact with a superconducting layer with a higher
''$T_{c}$'', superconductivity is predicted to be enhanced and
$T_{c}$ increases with reduced radius of curvature of the
grains\cite{montevecci}.

In this paper we review and refine the evidence for finite size
behavior in the specific heat and penetration depth data of
cuprate superconductors. Since $^{4}$He confined to uniform small
dimensions has been the subject of fundamental research for over
three decades, we sketch next the finite size scaling analysis of
the specific heat of \ $^{4}$He confined to cylindrical boxes
whose diameter and height are 1 $\mu $m\cite{kimball}. This
provides information on a system where confinement is made by
design and modifies its critical behavior, while other effects
such as disorder are not present. Close to the bulk transition
temperature $T_{c}$ the singular part of the specific heat adopts
the form
\begin{equation}
c_{b}\left( t\right) =c_{bs}\left( t\right) +B^{\pm },\text{
}c_{bs}\left( t\right) =\frac{A^{\pm }}{\alpha }\left| t\right|
^{-\alpha }.  \label{eq1}
\end{equation}
$t=1-T/T_{c}$ is the reduced temperature, $A^{\pm }$ the critical
amplitude and $\alpha $ the critical exponent of the specific
heat, while $B^{\pm }$ arises from the background, and $\pm
=sgn(t)$.

Since superconductors in the experimentally accessible critical
regime and $^{4}$He fall into the 3D-XY universality class, with
known critical exponents and critical amplitude combinations, we
take these properties for granted\cite{peliasetto}. They include
the exponents
\begin{equation}
\alpha =2-3\nu =-0.013,\ \nu =0.671,  \label{eq2}
\end{equation}
and the critical amplitude combinations
\begin{equation}
A^{\pm }V_{c}^{\pm }=\left( R^{\pm }\right) ^{3},V_{c}^{\pm
}=\left( \xi _{0}^{\pm }\right) ^{3},  \label{eq3}
\end{equation}
where
\begin{equation}
\frac{A^{+}}{A^{-}}=1.07,\ R^{-}=0.815,\text{ }R^{+}=0.361.
\label{eq4}
\end{equation}
$\xi _{0}^{\pm }$ is the critical amplitude of the correlation
length, diverging in the bulk system as $\xi ^{\pm }=\xi _{0}^{\pm
}\left| t\right| ^{-\nu }$, and $V_{c}^{\pm }=\left( \xi _{0}^{\pm
}\right) ^{3}$ is the correlation volume.

\begin{figure}[tbp]
\centering
\includegraphics[totalheight=6cm]{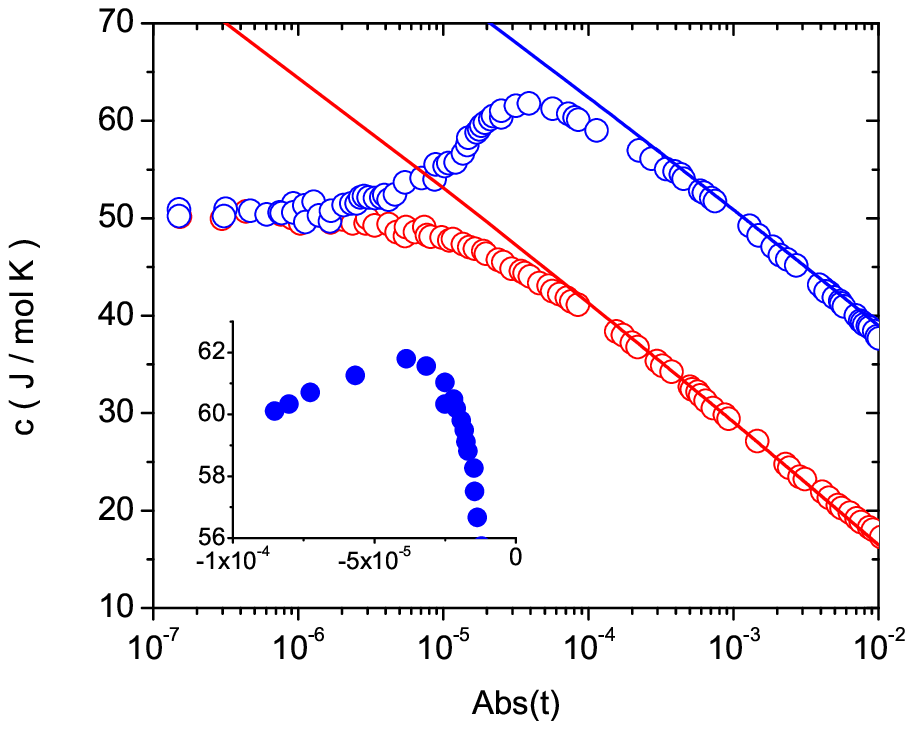}
\includegraphics[totalheight=6cm]{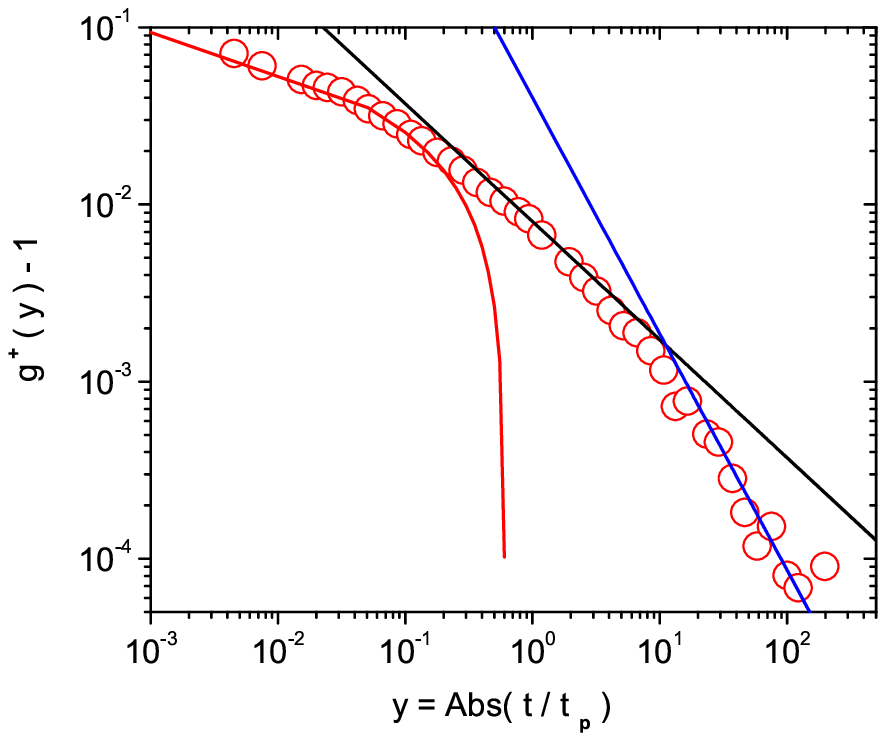}
\caption{(a) Specific heat $c$ of $^{4}$He confined to 1$\mu
m^{3}$ cylindrical boxes versus $\left| t\right| $compared to the
bulk data. The lines represent the bulk data and the open symbols
are the confined system. Taken from Kimball {\em et
al.}\protect\cite{kimball}. The insertion shows $c$ versus $t$
around $t_{p}\approx -3.9$ $10^{-5}$. (b) Corresponding finite
size scaling function $g^{+}\left( y\right) -1$ versus $y=\left|
t/t_{p}\right| $. The red curve indicates the limiting behavior
$g^{+}\left( \left| y\right| \rightarrow 0\right) =g_{0}^{+}\left|
y\right| ^{-\left| \alpha \right| }$ in terms of
$g^{+}=0.993\left| y\right| ^{-0.014}$. The black line,
$g^{+}\left( y\right) -1=0.008\left| y\right| ^{-2/3}$ uncovers
the surface and the blue one, $g^{+}\left( y\right) -1=0.04\left|
y\right| ^{-4/3}$ the edge contributions.} \label{fig1}
\end{figure}

Fig.\ref{fig1}a shows the specific heat of $^{4}$He confined to
1$\mu m^{3}$ cylindrical boxes and contrasts it to the unconfined
system, the solid lines \cite{kimball}. The lower branch
corresponds to $T>T_{c}$ and the upper to $T<T_{c}$. Notice at
large values of $\left| t\right| $, the data match the bulk data.
As one moves closer to the bulk transition, the confined systems
specific heat begins to systematically deviate from the bulk data.
This is predicted by finite-size scaling
theory\cite{privman,cardy}. As the system feels its finite size
when the correlation length $\xi $ becomes of the order of the
smallest confining length, the specific heat adopts the scaling
form, asserting that
\begin{equation}
\frac{c_{b}\left( t\right) -c\left( t,L\right) }{\left|
c_{bs}\left( t\right) \right| }=g^{\pm }\left( y\right) -1,\text{
\ }y=t/\left| t_{p}\right| ,\text{ \ }\left| t_{p}\right| =\left(
\xi _{0}^{-}/L\right) ^{1/\nu }.  \label{eq5}
\end{equation}
$g^{\pm }\left( y\right) $ is the finite size scaling function
which satisfies: $g^{-}(-1)=1$ at $T=T_{p}$, where $\xi ^{-}\left(
t_{p}\right) =\xi _{0}^{-}\left| t_{p}\right| ^{-\nu }=L$, so that
$y=y_{p}=-1$, while $g^{\pm }\left( y\rightarrow 0\right)
=g_{0}^{\pm }\left| y\right| ^{-\left| \alpha \right| }$ holds for
$\alpha <0$ and $g^{\pm }\left( \left| y\right| \rightarrow \infty
\right) =1$. $L$ is the smallest spatial length of the confined
system. $g^{+}\left( y\right) $, derived from the data of Kimball
{\em et al}.\cite{kimball}, is displayed in Fig.\ref{fig1}b. The
red curve indicates the 0D limiting behavior $g^{+}\left( \left|
y\right| \rightarrow 0\right) =g_{0}^{+}\left| y\right| ^{-\left|
\alpha \right| }$. In between these limits, where the correlation
length is smaller than $L$, so that $\left| t\right| \gtrsim
\left| t_{p}\right| $, the free energy density (and other
thermodynamic quantities) can be expanded in $D=3$
as\cite{privman}
\begin{equation}
f\left( t,L\right) -f_{b}\left( t\right) =\frac{1}{L}f_{s}\left( t\right) +%
\frac{1}{L^{2}}f_{e}\left( t\right) +\frac{1}{L^{3}}f_{c}\left(
t\right). \label{eq6}
\end{equation}
The contributions $f_{s}$, $f_{e}$ and $f_{c}$ can be attributed
to the surfaces, edges and corners of the confinement. It is
expected that each one of these terms contribute in a limited
region of the scaling variable. Since $f_{s}$ and $f_{e}$ scale as
$\xi ^{-2}$ and $\xi ^{-1}$, their contribution
to the specific heat $c\propto -\partial ^{2}f/\partial t^{2}$ scales as $%
\left| t\right| ^{-\alpha -\nu }$ and $\left| t\right| ^{-\alpha
-2\nu }$, respectively. Hence the surface term should behave as
$\left| t/t_{p}\right| ^{-\nu }$ and the edge term as $\left|
t/t_{p}\right| ^{-2\nu }$ in the scaling form, Eq. (\ref{eq5}). A
glance to Fig.\ref{fig1}b shows that as one proceeds to smaller
values of the scaling variable $y=\left|
t/t_{p}\right| $ there is clear evidence for crossovers from the edges, $%
\propto \left| t/t_{p}\right| ^{-4/3}$, to the surface, $\propto
\left| t/t_{p}\right| ^{-2/3}$, contribution and finally to the
asymptotic behavior, $g^{+}\left( \left| y\right| \rightarrow
0\right) =g_{0}^{+}\left| y\right| ^{-\left| \alpha \right| }$.
Thus, the finite size scaling function does not uncover the
crossover from 3D to 0D behavior only, but allows to identify
surface and edge contributions, as well.

We have seen that $^{4}$He confined to 1$\mu m^{3}$ cylindrical
boxes crosses over from a $3D$ to a $0D$ behavior near the bulk
transition. This has a marked effect on the specific heat as seen
by the pronounced rounding of the peak and its shift to a
temperature lower than the transition of the bulk system.
Furthermore, the region in between the $3D$ to $0D$ crossover
uncovers the contributions from the surface and the edges of the
cylindrical boxes. Since disorder is not present, the
modifications of the critical behavior are entirely due to the
confinement. Analogous behavior in cuprate superconductors would
imply that the bulk breaks into homogeneous superconducting grains
with rather unique extent, embedded in a nonsuperconducting
matrix.

\begin{figure}[tbp]
\centering
\includegraphics[totalheight=6cm]{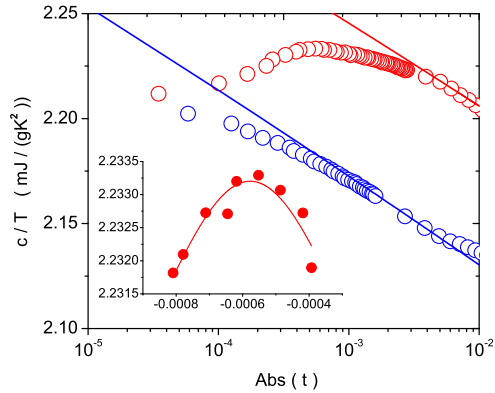}
\caption{Specific heat coefficient $c/T$ of the high quality YBa$_{2}$Cu$_{3}$O%
$_{7-\delta }$ single crystal (sample UBC2) versus $\left|
t\right| $ taken from Charalambous {\em et
al}.\protect\cite{chara}. The solid lines represent the expected
behavior of a fictitious homogeneous bulk sample according to
Eq.(\ref{eq1}). The insertion shows $c/T$ versus $t$ around
$t_{p}\approx -5.8$ $10^{-4}$.} \label{fig2}
\end{figure}

In Fig.\ref{fig2} we show the specific heat data of Charalambous
{\em et al}. \cite{chara} of a high quality
YBa$_{2}$Cu$_{3}$O$_{7-\delta }$ single crystal in terms of the
specific heat coefficient $c/T$ versus $\left| t\right| $.
Comparing Figs.\ref{fig1}a and \ref{fig2} we observe: (i) in
YBa$_{2}$Cu$_{3}$O$_{7-\delta }$ the rounding of the transition
sets in at much higher reduced temperatures; (ii) otherwise the
rounding is remarkably analogous to that of $^{4}$He confined to
1$\mu m^{3}$ cylindrical boxes. Noting that in $^{4}$He and
optimally doped YBa$_{2}$Cu$_{3}$O$_{7-\delta }$ the critical
amplitude $\xi _{0}^{-}$ are comparable, the relation $\left|
t_{p}\right| =\left( \xi _{0}^{-}/L\right) ^{1/\nu }$
(Eq.\ref{eq2}) indicates that the difference in the onset of the
rounding is due to a limiting length scale, considerably smaller
than the 1$\mu m$ in $^{4}$He.

\begin{figure}[tbp]
\centering
\includegraphics[totalheight=6cm]{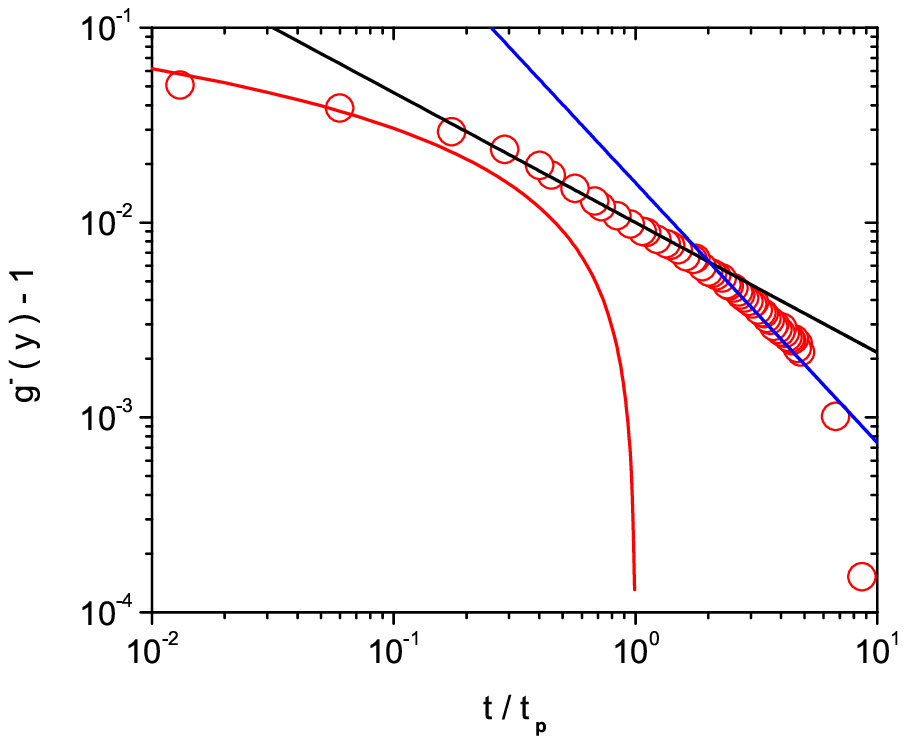}
\includegraphics[totalheight=6cm]{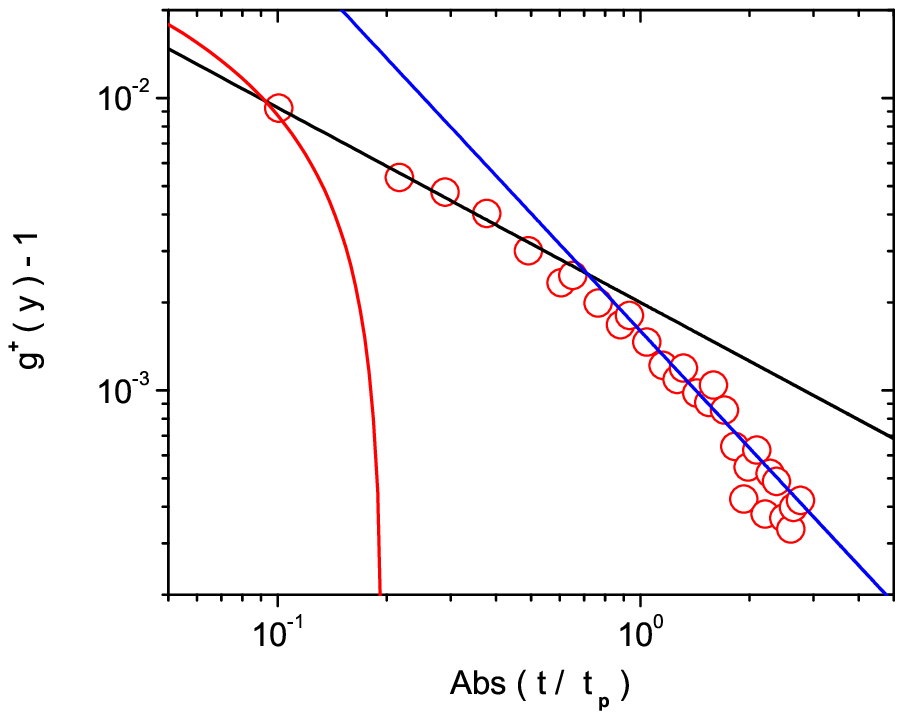}
\caption{(a) $g^{-}\left( y\right) -1$ versus $y=t/t_{p}$ derived
from the
data shown in Fig.\ref{fig2}. The red curve indicates the limiting behavior $%
g^{-}\left( \left| y\right| \rightarrow 0\right) =g_{0}^{-}\left|
y\right| ^{-\left| \alpha \right| }$ in terms of $g^{+}=\left|
y\right| ^{-0.013}$. The black line, $g^{-}\left( y\right)
-1=0.01\left| y\right| ^{-2/3}$ uncovers the surface and the blue
one, $g^{-}\left( y\right) -1=0.016\left| y\right| ^{-4/3}$ the
edge contributions.(b) $g^{-}\left( y\right) -1$ versus $y=\left|
t/t_{p}\right| $ derived from the data shown in Fig.\ref {fig2}.
The red curve indicates the limiting $0$D behavior, $g^{+}\left(
\left| y\right| \rightarrow 0\right) =\left| y\right| ^{-0.013}$.
The black line, $g^{+}\left( y\right) -1=0.002\left| y\right|
^{-2/3}$ corresponds to the surface and the blue one, $g^{+}\left(
y\right) -1=0.0016\left| y\right| ^{-4/3}$, to the edge
contributions.} \label{fig3}
\end{figure}

To substantiate and refine the consistency with a finite size
effect further we displayed in Fig.\ref{fig3} the scaling
functions $g^{-}\left( y\right) -1 $ and $g^{+}\left( y\right)
-1$. As one proceeds to smaller values of the scaling variable
$y=\left| t/t_{p}\right| $ there is, in analogy to $^{4}$He
confined to 1$\mu m^{3}$ cylindrical boxes (Fig.\ref{fig1}b),
clear evidence for crossovers from the edge, $\propto $ $\left|
t/t_{p}\right| ^{-4/3}$, to the surface, $\propto \left|
t/t_{p}\right| ^{-2/3}$, contributions and finally to the 0D
behavior, $g^{+}\left( \left| y\right| \rightarrow 0\right)
=g_{0}^{+}\left| y\right| ^{-\left| \alpha \right| }$.

\begin{figure}[tbp]
\centering
\includegraphics[totalheight=6cm]{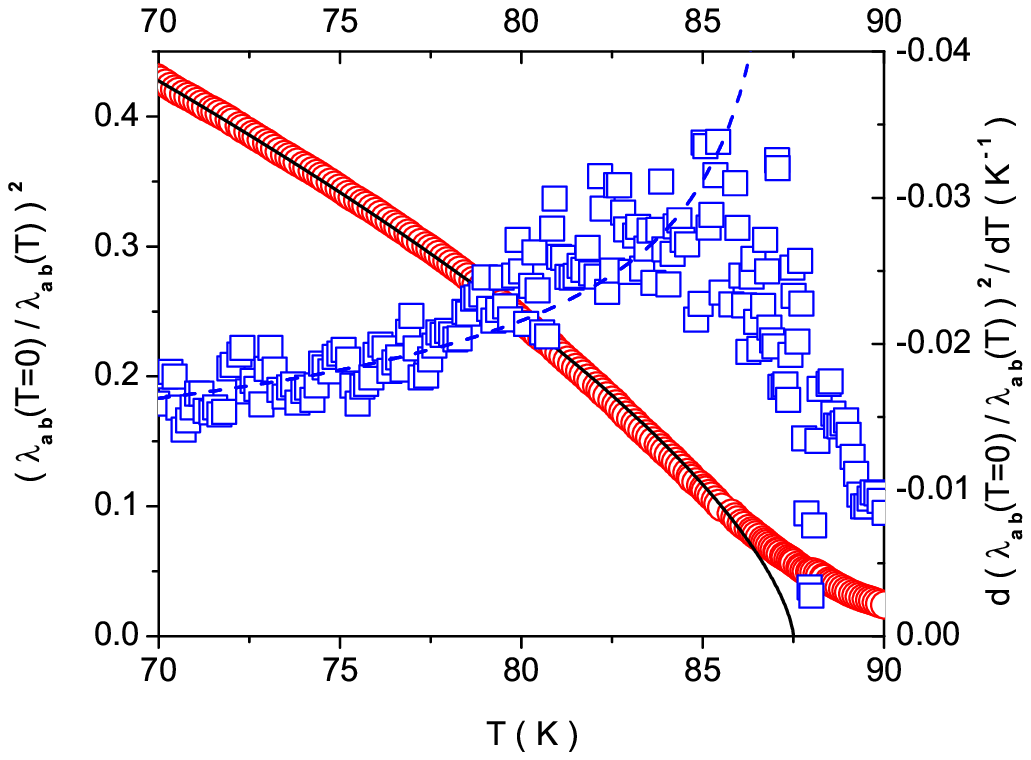}
\includegraphics[totalheight=6cm]{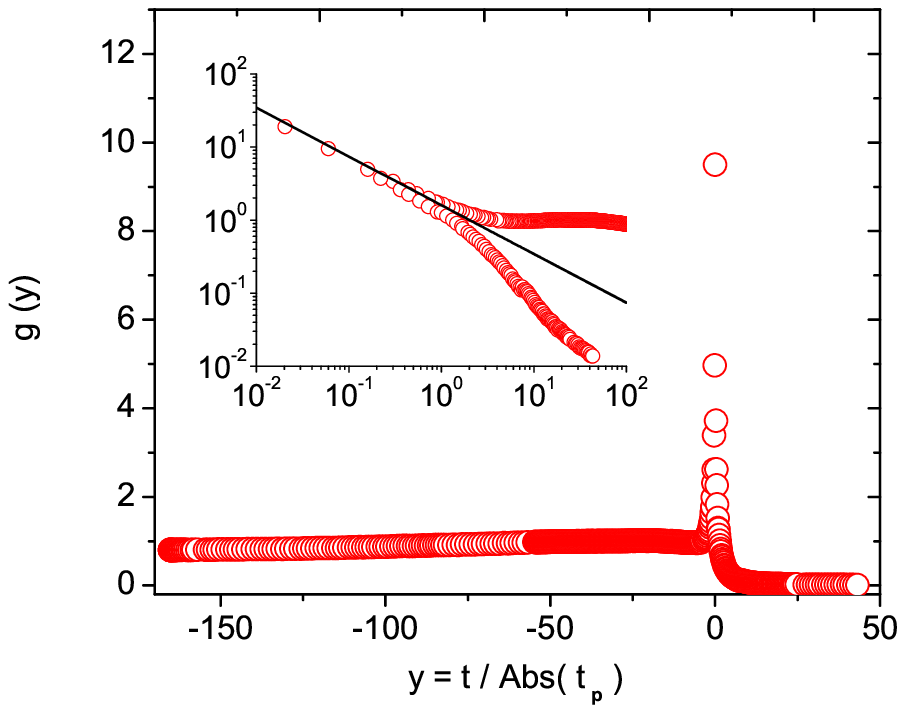}
\caption{(a)Microwave surface impedance data for $\lambda
_{ab}^{2}\left( T=0\right) /\lambda _{ab}^{2}\left( T\right) $
($\bigcirc $)\ and $d\left( \lambda _{ab}^{2}\left( T=0\right)
/\lambda _{ab}^{2}\left( T\right) \right)
/dT$ ($\square $) versus $T$ of a high-quality Bi$_{2}$Sr$_{2}$CaCu$_{2}$O$%
_{8+\delta }$ single crystal taken from Jacobs {\em et al.}
\protect\cite{jacobs}. The solid line is $\lambda _{ab}^{2}\left(
0\right) /\lambda _{ab}^{2}\left( T\right) =1.2\left(
1-T/T_{c}\right) ^{2/3}$ with $T_{c}=87.5$K and the dashed line
its derivative indicating the leading critical behavior of the
fictitious
homogeneous system. The rounded transition exhibits an inflection point at $%
T_{p}\approx 87$K, where $d\left( \lambda _{ab}^{2}\left( 0\right)
/\lambda
_{ab}^{2}\left( T\right) \right) /dT$ is maximum. (b) Scaling function $%
g\left( y\right) =\left( \lambda _{0ab}/\lambda _{ab}\left(
T\right) \right) ^{2}\left| t\right| ^{-\nu }$ versus $y=t/\left|
t_{p}\right| $ for the data shown in Fig.\ref{fig4}a. The solid
line in the insertion is Eq.(\ref{eq10}) with $g_{0}=1.6$. It
indicates the flow to the 0D behavior.} \label{fig4}
\end{figure}

When this finite size scenario holds true, the London penetration
depth should not only exhibit a rounded transition, consistent
with a finite size effect\cite{tsrkhk,varenna,tsdc,bled,rktshkp},
but uncover the surface and edge contributions, as well.
Considering again the 3D-XY critical point, extended to the
anisotropic case, the penetration depths and transverse
correlation lengths in directions $i$ and $j$ are universally
related by\cite {book,hohenberg}
\begin{equation}
\frac{1}{\lambda _{i}\left( T\right) \lambda _{j}\left( T\right) }=\frac{%
16\pi ^{3}k_{B}T}{\Phi _{0}^{2}\sqrt{\xi _{i}^{t}\left( T\right)
\xi _{j}^{t}\left( T\right) }}.  \label{eq7}
\end{equation}
A limiting length scale $L_{i}$ in direction $i$ implies that $\xi
_{i}^{t}\xi _{j}^{t}$ does not diverge but is bounded by $\xi
_{i}^{t}\xi _{j}^{t}=\left( \xi _{k}^{-}\right) ^{2}\leq
L_{k}^{2}$, where $i\neq j\neq k $. $\xi _{i}^{t}$ denotes the
transverse and $\xi _{k}^{-}=\sqrt{\xi _{i}^{t}\xi _{j}^{t}}$ the
corresponding real space correlation length. The resulting finite
size effect has been established in a variety of cuprates
\cite{tsrkhk,varenna,tsdc,bled,rktshkp}. Here we refine these
studies to identify surface and edge contributions. As first
example we consider the high-quality
Bi$_{2}$Sr$_{2}$CaCu$_{2}$O$_{8+\delta }$ single crystal data of
Jacobs {\em et al.} \cite{jacobs} shown in Fig.\ref{fig4}. The
solid curve indicates the leading critical behavior of the
fictitious  homogeneous system with $T_{c}=87.5$K, while the
rounded transition points to a finite size effect. A
characteristic feature of a finite size effect in $1/\lambda
_{ab}^{2}\left( T\right) $ is the occurrence of an inflection
point giving rise to an extremum in $d\left( \lambda
_{ab}^{2}\left( T=0\right) /\lambda _{ab}^{2}\left( T\right)
\right) /dT$ at $T_{p}$. Here Eq.(\ref{eq7}) reduces to
\begin{equation}
\frac{1}{\lambda _{ab}^{2}\left( T_{p}\right) }\approx
\frac{1}{\lambda _{a}\left( T_{p}\right) \lambda _{b}\left(
T_{p}\right) }=\frac{16\pi ^{3}k_{B}T_{p}}{\Phi _{0}^{2}L_{c}}.
\label{eq8}
\end{equation}
The data shown in Fig.\ref{fig4} exhibits in $d\left( \lambda
_{ab}^{2}\left( T=0\right) /\lambda _{ab}^{2}\left( T\right)
\right) /dT$ at
$T_{p}\approx 87$K an extremum and with that an inflection point in $%
1/\lambda _{ab}^{2}\left( T\right) $. With $\lambda _{ab}\left(
T=0\right) =1800$\AA\ as obtained from $\mu $SR measurements
\cite{leem} and $\lambda _{ab}^{2}\left( T=0\right) /\lambda
_{ab}^{2}\left( T_{P}\right) =0.066$ we find $L_{c}\approx 68$\AA.

\begin{figure}[tbp]
\centering
\includegraphics[totalheight=6cm]{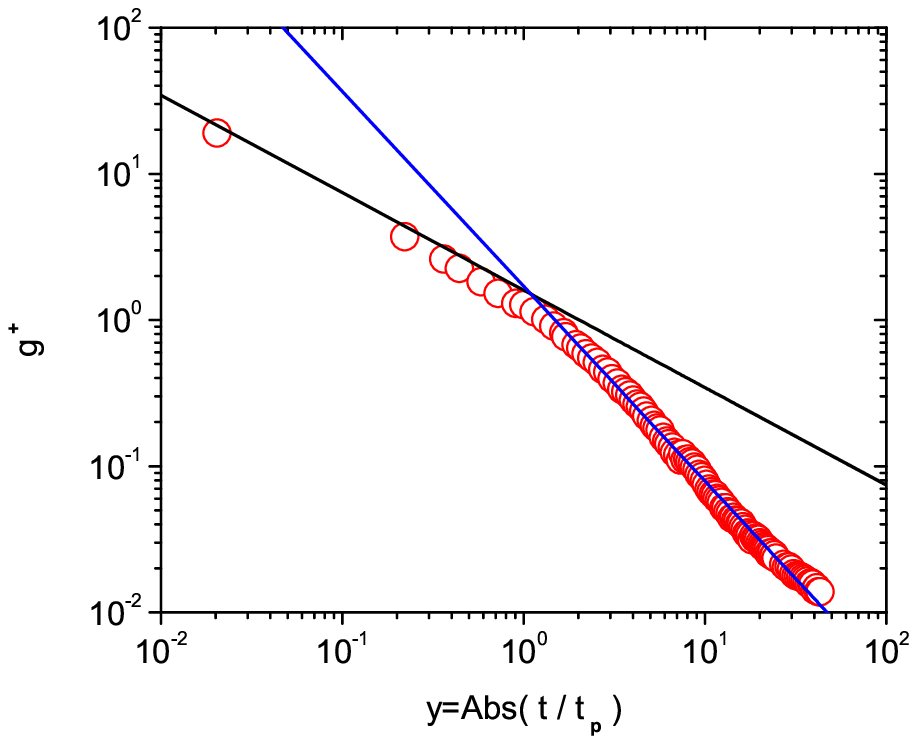}
\includegraphics[totalheight=6cm]{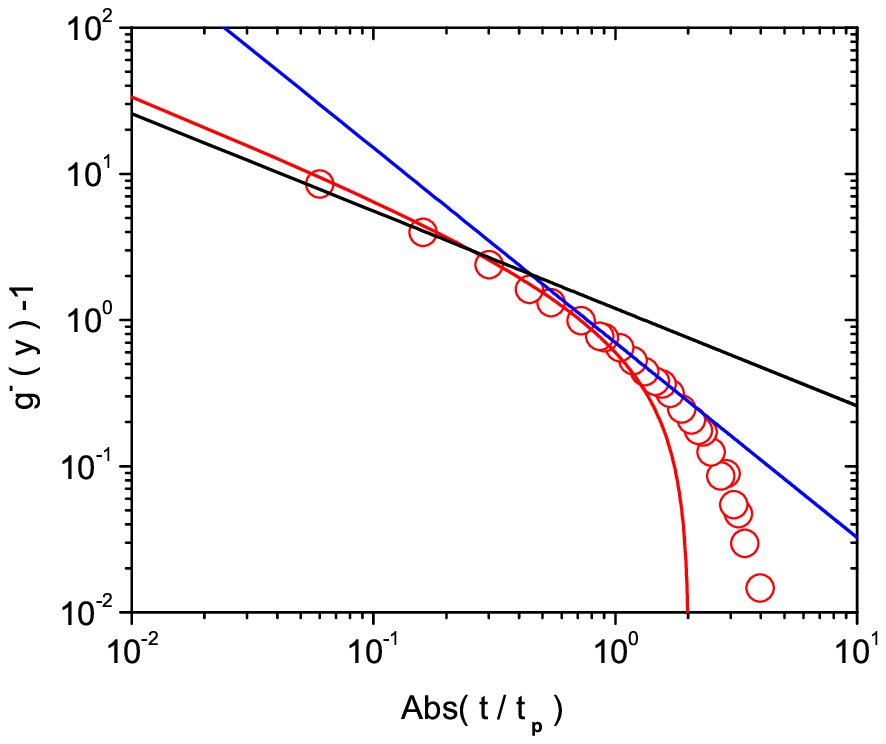}
\caption{(a) Scaling function $g^{+}\left( y\right) $ versus
$y=\left| t/t_{p}\right| $ derived from the data shown in
Fig.\ref{fig4}. The blue line, $g^{+}\left( y\right) =1.7y^{-4/3}$
corresponds to the edge and the black one, $g^{+}\left( y\right)
=1.6y^{-2/3}$, to the surface contribution. (b) $g^{-}\left(
y\right) -1$ versus $y=\left| t/t_{p}\right| $ derived from the
data shown in Fig.\ref{fig4}. The blue line, $g^{-}\left( y\right)
-1=0.7y^{-4/3}$, indicates the edge, the black one, $g^{-}\left(
y\right) -1=1.2y^{-2/3}$, the surface contribution and the red
curve indicates the flow to the $0$D behavior, $g^{-}\left(
y\right) =1.6y^{-2/3}$.} \label{fig5}
\end{figure}

To strengthen the evidence for a finite size effect we explore
next the consistency of the data with the corresponding finite
size scaling function. In the present case it is defined as
\begin{equation}
\left( \frac{\lambda _{0ab}}{\lambda _{ab}\left( T,L_{c}\right)
}\right) ^{2}\left| t\right| ^{-\nu }=g^{\pm }\left( y\right) ,\ \
y=sign\left( t\right) \left| \frac{t}{t_{p}}\right| .  \label{eq9}
\end{equation}
For $t$ small and $L_{c}\rightarrow \infty $, so that $\pm
y\rightarrow \infty $ it should tend to $g^{-}\left( y\rightarrow
-\infty \right) =1$ and $\ g^{+}\left( y\rightarrow \infty \right)
=0$, respectively, while for $t=0$ and $L\neq 0$ it diverges as
\begin{equation}
g^{\pm }\left( y\rightarrow 0\right) =g_{0}^{\pm }\left| y\right|
^{-\nu }. \label{eq10}
\end{equation}
In this 0D limit, $\left( \lambda _{0ab}/\lambda _{ab}\left(
T_{c},L_{c}\right) \right) ^{2}=g_{0}^{-}\left| t_{p}\right| ^{\nu
}=g_{0}^{-}\xi _{0c}^{-}/L_{c}$. Moreover at $t_{p}$, $y=-1$ and
$d\left( \lambda _{0ab}/\lambda _{ab}\left( T,L\right) \right)
^{2}/dt=0$. The scaling function shown in Fig.\ref{fig4}b, derived
from the data shown in Fig.\ref{fig4}a, is apparently fully
consistent with the limiting behavior of the finite size scaling
function and uncovers, in analogy $^{4}$He, confined to
cylindrical boxes, a 3D to 0D crossover. On the other hand, in
analogy to the specific heat, refined information may be attained
in between these limits. Here the correlation length is smaller
than $L_{c}$, $\left| t\right| \gtrsim \left| t_{p}\right| $, and
the contributions from the surface and edges should become
observable. Since $f_{s}$ and $f_{e}$ scale as $1/\left( L\xi
^{2}\right) $ and $1/\left( L^{2}\xi \right) $, their
contributions to $1/\lambda ^{2}\propto f$ $\xi ^{2}$ scale as
$1/L$ and $\xi /L$, respectively. Hence the surface term should
behave as $\left| t/t_{p}\right| ^{-\nu }$ and the edge term as
$\left| t/t_{p}\right| ^{-2\nu }$ in the scaling form, Eq.
(\ref{eq9}). A glance to Fig.\ref{fig5} shows that as one proceeds
to smaller values of the scaling variable $y=\left| t/t_{p}\right|
$ there is indeed evidence for crossovers from the edges, $\propto
\left| t/t_{p}\right| ^{-4/3}$, to the surface, $\propto \left|
t/t_{p}\right| ^{-2/3}$, contributions and finally to the 0D
behavior, $g^{+}\left( \left| y\right| \rightarrow 0\right)
=g_{0}^{+}\left| y\right| ^{-2/3}$.

\_

\begin{figure}[tbp]
\centering
\includegraphics[totalheight=6cm]{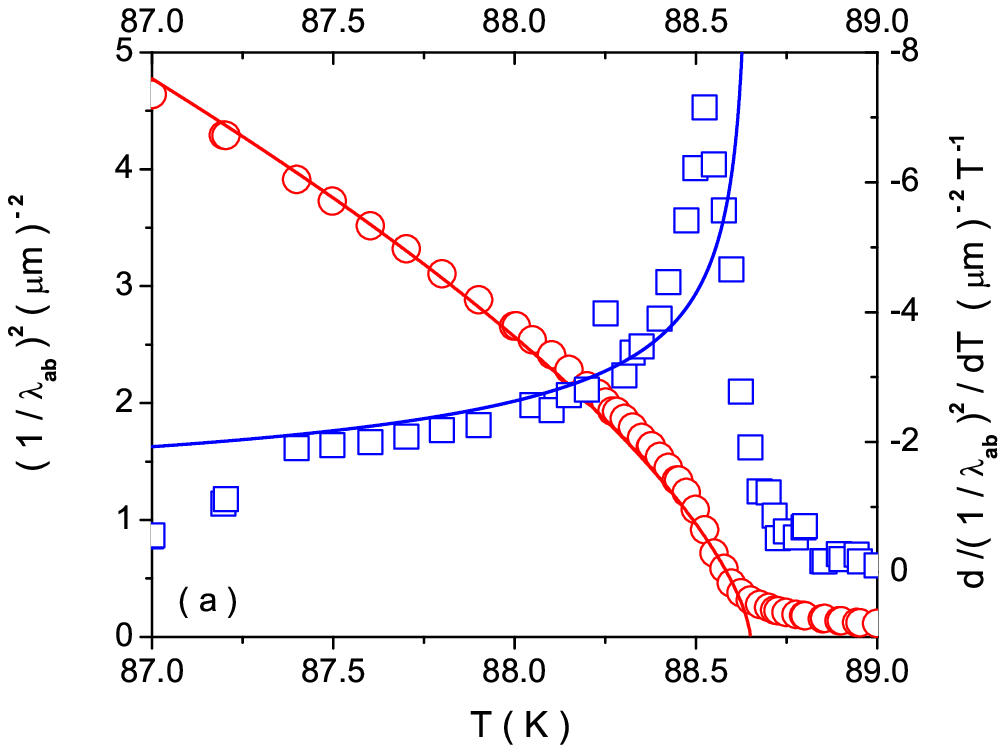}
\includegraphics[totalheight=6cm]{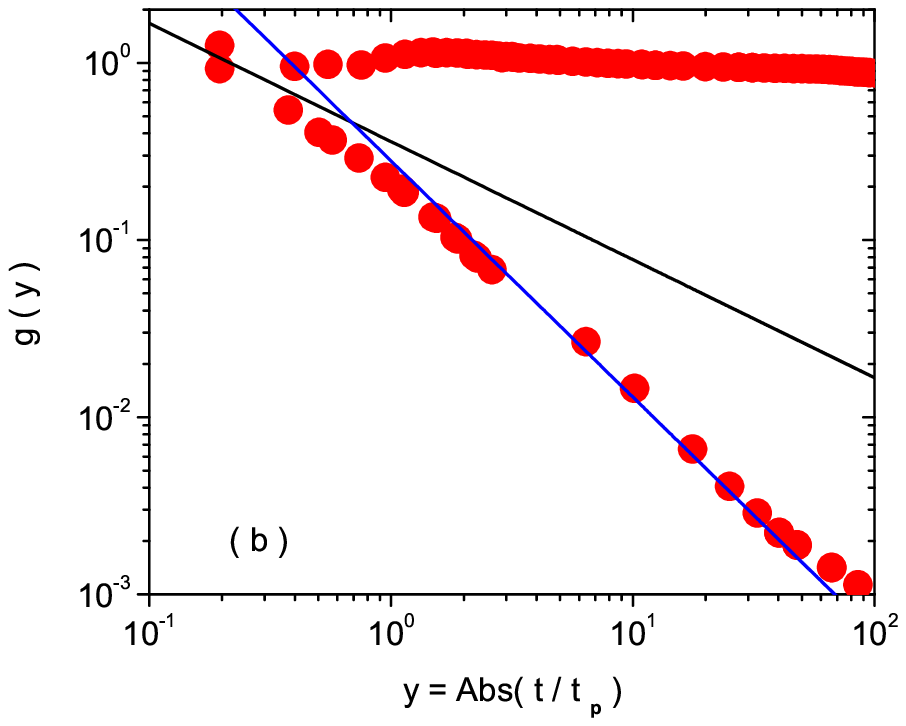}
\caption{(a) In-plane penetration depth data for $1/\lambda
_{ab}^{2}\left( T\right) $ ($\bigcirc $)\ and $d\left( 1/\lambda
_{ab}^{2}\left( T\right) \right) /dT$ ($\square $) versus $T$ of a
high-quality YBa$_{2}$Cu$_{3}$O$_{6.95}$ single crystal taken from
Kamal {\em et al.} \protect\cite{kamal}. The solid line is
$1/\lambda _{ab}^{2}\left( T\right) =68\left( 1-T/T_{c}\right)
^{2/3}$ with $T_{c}=88.65$K and the dashed line its derivative
indicating the leading critical behavior of the fictitious
homogeneous bulk system. The rounded transition exhibits an
inflection point at $T_{p}\approx 88.52$K, where $d\left(
1/\lambda _{ab}^{2}\left( T\right) \right) /dT$ is maximum. (b)
Scaling function $g\left( y\right) =\left( \lambda _{0ab}/\lambda
_{ab}\left( T\right) \right) ^{2}\left| t\right| ^{-\nu }$ versus
$y=t/\left| t_{p}\right| $ for the data shown in Fig.\ref{fig4}a.
The black line, $g^{+}\left( y\right) =0.36y^{-2/3}$, indicates
the surface and the blue one, $g^{+}\left( y\right)
=0.28y^{-4/3}$, the edge contributions above $T_{c}$.}
\label{fig6}
\end{figure}

As second example we treat the in-plane penetration depth data of
Kamal {\em et al.} \cite{kamal} for a high-quality
YBa$_{2}$Cu$_{3}$O$_{6.95}$ single crystal. In Fig.\ref{fig6}a we
displayed the data in terms of $1/\lambda _{ab}^{2}\left( T\right)
$ and $d\left( 1/\lambda _{ab}^{2}\left( T\right) \right) /dT$
versus $T$. The red and blue curves indicate the leading critical
behavior of the fictitious homogeneous system with $T_{c}=88.65$K,
while the rounded transition points to a finite size effect. The
occurrence of an extremum in $d\left( 1/\lambda _{ab}^{2}\left(
T\right) \right) /dT$ at $T_{p}\approx 88.52$K is a characteristic
feature of a finite size effect in $1/\lambda _{ab}^{2}\left(
T\right) $, giving rise to an inflection point in $1/\lambda
_{ab}^{2}\left( T\right) $ at $T_{p}$. Using $1/\lambda
_{ab}^{2}\left( T_{p}\right) =0.918$ ($\mu $m)$^{-2}$, we deduce
with the aid of Eq.(\ref{eq8}) for the limiting length scale along
the c-axis the estimate $L_{c}=154$\AA . The associated scaling
function $g\left( y\right) $, defined in Eq.(\ref{eq9}), is shown
in Fig.\ref{fig6}b. As one proceeds to smaller values of the
scaling variable $y$ there is in the lower branch ($T>T_{c}$)
clear evidence for a crossover from the edge, $\propto $ $\left|
t/t_{p}\right| ^{-4/3}$, to the surface, $\propto \left|
t/t_{p}\right| ^{-2/3}$, contributions.

We reviewed and refined the finite size scaling analysis of
specific heat and London penetration depths data of cuprate
superconductors and compared them to the analysis of specific heat
measurements near the superfluid transition of $^{4}$He confined
to 1$\mu $m$^{3}$ cylindrical boxes. The $^{4}$He system provides
information on a system where confinement is made deliberately. It
modifies its critical behavior while other effects such as
disorder are not present. We have seen that this system crosses
from 3D to 0D behavior near the transition. As a result of this
crossover, the specific heat peak exhibits a pronounced rounding
and the maximum shifts to a temperature lower than the transition
temperature of the bulk system. The region in between the 3D to 0D
crossover uncovered contributions from the surface and the edges
of the cylindrical boxes. The analysis of the specific heat and
London penetration depth of high quality
YBa$_{2}$Cu$_{3}$O$_{7-\delta }$ and
Bi$_{2}$Sr$_{2}$CaCu$_{2}$O$_{8+\delta }$ single crystals
uncovered essentially the same crossover phenomena, including
evidence for surface and edge contributions. No new critical
behavior emerges, but 3D-XY scaling is achieved with the variable
$\xi /L$, where $L$ is the smallest confining dimension. This
implies that the bulk samples break into nearly homogeneous
superconducting grains of rather unique extent and with that
granular superconductivity.

\bigskip
\acknowledgments The author is grateful to D. Di Castro, R.
Khasanov, H. Keller, K.A. M\"{u}ller and J. Roos for very useful
comments and suggestions on the subject matter.

\end{document}